\documentclass[11pt]{article}

\usepackage{hyperref}
\usepackage{amsmath,amsfonts,amssymb}
\hypersetup{dvips,dvipdfm,colorlinks=true,urlcolor=magenta,filecolor=magenta,linktoc=page,citecolor=red,linkcolor=blue,bookmarks=true}
\usepackage{graphicx,epstopdf}
\usepackage{multirow,array}
\begin{document}
	\title{\textbf{Cosmological Wormhole:  An analytical description}}
	\author{Subenoy Chakraborty \footnote{\url{schakraborty.math@gmail.com} (corresponding author)}~~and~~
		Madhukrishna Chakraborty\footnote{\url{chakmadhu1997@gmail.com}}
		\\ \textit{Department of Mathematics, Jadavpur University, Kolkata - 700032, India}}
	\date{}
	\maketitle
	\begin{abstract}
	An attempt has been made to have an analytical description for possible traversable wormhole in non-static spherically symmetric space-time supported by anisotropic fluid. Both trivial and non-trivial choices of the red-shift function result in identical WH configuration and it is possible to have emergent scenario for evolution of the background space-time in both the cases.
	\end{abstract}
	\maketitle
	\section{Introduction}
Wormhole configuration is an interesting and popular topic of investigation in relativistic astrophysics. It indicates rapid interstellar travel and was formalized long back in 1988 by Morris and Throne \cite{Morris:1988cz}. The notion of traversability means a bi-directional passage through the throat which acts as the passage or connection between distant regions of the same universe or distant universes  \cite{Morris:1988cz}-\cite{Visser:1998ua}. A general speculation about WH geometry is to have a long traveler through this WH having super-luminal travel for global space-time topology, without surpassing the speed of light (locally). As a consequence, the hypothetical notion of time machine \cite{Visser:1995cc}, \cite{Visser:1998ua} has analogy with WH geometry.

 The primary condition for Wormholes (WH) to be traversable \cite{Lobo:2007zb}, \cite{Morris:1988tu} is to have a red-shift function without horizons or a given asymptotic form for both the red-shift and the shape function is desirable \cite{Kim:1997jf}, \cite{Halder:2019urh}. The common geometric notion of traversability is known as the flare-out condition which in general relativity indicates exotic matter (violating the null energy condition) threading the throat \cite{Morris:1988cz}, \cite{Visser:1995cc}, \cite{Hochberg:1997wp}-\cite{Fewster:2005gp}. However, the amount of exotic matter (around the throat) can be made infinitesimally small (i.e, violation of the averaged null energy condition (NEC)) \cite{Visser:2003yf} with suitable choice of the WH geometry but at the cost of large stresses at the throat \cite{Eiroa:2005pc}, \cite{Zaslavskii:2007gu} . Further, in asymptotically flat space-time, this violation of NEC is a consequence of the topological censorship \cite{Friedman:1993ty}. Hence, it is reasonably speculated that WH configuration may be formed by quantum effects, violating the energy conditions. 
 
 The common way of constructing (theoretical) WH geometry is to assume a priori the desired form of both the red-shift function and the shape function and the possible matter field is determined via Einstein field equations. Alternatively, by choosing realistic fluid for the matter part one may determine the the red-shift function and shape function  via Einstein field equations \cite{Halder:2019urh}. In the first approach the matter density and pressure are obtained in algebraic form. They may not be always realistic while for the other option traversability of the WH or consistency of the field equations often demands the phantom nature of the fluid component \cite{Visser:1998ua}. On the other hand, neither the dynamical or relativistic evolving WHs \cite{Cataldo:2008pm}, \cite{Cataldo:2013ala} are so common as the static ones nor do they have a clear picture. Hochberg, Visser \cite{Hochberg:1998ii}, \cite{Hochberg:1998ha} and independently Hayward \cite{Hayward:1998pp} are pioneers for studying dynamical WHs, by choosing quasi local definition of WH throat in a dynamical space-time. In fact, dynamical wormhole is a different kind of trapping horizon \cite{Hayward:1998pp} with matter violating NEC. Alternatively, (CWH) (i.e, cosmological WHs) having asymptotically Friedmann universe has as big-bang singularity. They do not violate NEC, rather the dominant energy condition is satisfied. The basic geometric difference between these two types of CWHs is that in the former type the WH throat is a $2D$ surface of non-vanishing minimal area on a null hyper-surface while for the second type due to initial singularity the WH throat exists on a space-like hyper-surface. Here there is no trapping horizon, rather the space-time being everywhere trapped \cite{Maeda:2009tk}. Usually, a two fluid system \cite{Cataldo:2012pw}, \cite{Pan:2013rha} is considered as the matter content for CWH configuration and it is very much relevant in the context of present day cosmology \cite{Pan:2013rha}, \cite{Cataldo:2008ku}, \cite{Dutta:2024luw}. The motivation of studying cosmological wormhole is to examine whether non-exotic matter may lead to WH traversibility unlike static and inflating WH models. In this context, the present work is an extension of a renowned work on CWH geometry \cite{Kim:1995xf}. A general prescription for the nature of matter parts has been considered and the general results have been presented in the form of lemmas. Also, the role of red-shift function has been examined in detail. The organization of the paper is as follows: Section 2 deals with an analytical description of the CWH configuration in inhomogeneous FLRW space-time. WH solutions and corresponding cosmic evolution have been given in Section 3 for zero red-shift function while wormhole solution with non-zero red-shift function has been presented in Section 4. The paper ends with a brief discussion and concluding remarks in Section 5.
\section{A brief description of CWH in inhomogeneous FLRW space-time}
The space-time geometry for CWH configuration is analogous to the inhomogeneous FLRW space-time and the line element for the same is given by \cite{Kim:1995xf}
\begin{equation}
	ds^{2}=-e^{2\phi(r,t)}dt^{2}+a^{2}(t)\left(\frac{dr^{2}}{1-\frac{b(r)}{r}-Kr^{2}}+r^{2}d\Omega_{2}^{2}\right)
\end{equation} where $a(t)$ is the scale factor, $b(r)$ is the shape function, $\phi(r,t)$ is the red-shift function, $K=0,\pm 1$ is the curvature index scalar and $d\Omega_{2}^{2}=d\theta^{2}+\sin^{2}\theta d\Phi^{2}$ is the metric on unit 2-sphere. The geometry will change accordingly as follows:
\begin{enumerate}
	\item Inhomogeneous FLRW space-time: $\phi=0$, $\dfrac{b(r)}{r}+Kr^{2}=C(r)$
	\item Homogeneous FLRW space-time: $\phi=0$, $a(t)=a_{0}$, $b(r)=0$
	\item Static spherically symmetric WH configuration: $\phi(r,t)=\phi_{0}(r),~a(t)=a_{0},~K=0$.
\end{enumerate}
The matter component is divided into two parts namely,
\begin{enumerate}
	\item Homogeneous, isotropic but dissipative in nature (Fluid I)
	\item Inhomogeneous and anisotropic in nature (Fluid II)
\end{enumerate}
The explicit form of the energy momentum tensors are
\begin{eqnarray}
	T_{\mu\nu}^{1}=(\rho_1+p_{1}+\Pi)u_{\mu}u_{\nu}+(p_{1}+\Pi)g_{\mu\nu}\\
	T_{\mu\nu}^{2}=(\rho_{2}+p_{t})v_{\mu}v_{\nu}+p_{t}g_{\mu\nu}+(p_{r}-p_{t})\dot{\chi_{\mu}}\chi_{\nu}
\end{eqnarray}  where $\rho_{1}=\rho_{1}(t)$, $p_{1}=p_{1}(t)$ and $\Pi=\Pi(t)$ are the energy density, thermodynamic pressure and dissipative pressure of Fluid I while $\rho_{2}=\rho_{2}(r,t)$, $p_{r}=p_{r}(r,t)$, $p_{t}=p_{t}(r,t)$ are the energy density, radial and transverse pressures of Fluid II. Usually, in CWHs the component of one of the components of fluid depends only on the radial coordinate and the other fluid component depends on the time coordinate. But, here we have chosen a general prescription where one of the fluid components depends on both the radial and temporal coordinate. Further, instead of perfect fluid, fluid I has some dissipative pressure component, while fluid II is anisotropic in nature.  So the matter conservation equation for the above non-interacting two fluids has the explicit form:
\begin{eqnarray}
	\dfrac{\partial \rho_{1}}{\partial t}+3H(\rho_{1}+p_{1}+\Pi)=0\label{eq1*}\\
	\dfrac{\partial \rho_{2}}{\partial t}+H(3\rho_{2}+p_{r}+2p_{t})=0\label{eq2*}
\end{eqnarray} and the Relativistic Euler equation is given by
\begin{equation}
	\dfrac{\partial p_{r}}{\partial r}=\dfrac{2}{r}(p_{t}-p_{r})\label{eq00}
\end{equation} Here, $u^{\mu}$, $v^{\mu}$ are unit time-like vectors i.e, $u_{\mu}u^{\mu}=v_{\mu}v^{\mu}=-1$ while $\chi^{\mu}$ is a unit space-like vector i.e, $\chi_{\mu}\chi^{\mu}=1$ and $v_\mu\chi^{\mu}=0$. Further, the explicit form of Einstein field equations: $G_{\mu\nu}=-\kappa(T_{\mu\nu}^{1}+T_{\mu\nu}^{2})$ are given by \cite{Cataldo:2012pw}
\begin{eqnarray}
	3e^{-2\phi(r,t)}H^{2}+\dfrac{b'}{a^{2}r^{2}}+\dfrac{3K}{a^{2}}=\kappa(\rho_{1}+\rho_{2})\label{eq4}\\
	-e^{-2\phi(r,t)}(2\dot{H}+3H^{2})+\dfrac{K}{a^{2}}-\dfrac{b}{a^{2}r^{3}}+2e^{-2\phi}H\dfrac{\partial \phi}{\partial t}+
	\dfrac{2}{a^{2}r^{2}}(r-b)\dfrac{\partial \phi}{\partial r}=\kappa(p_{1}+p_{r}+\Pi)\label{eq5}\\
	-e^{-2\phi}(2\dot{H}+3H^{2})+\dfrac{K}{a^{2}}+\dfrac{b-rb'}{2a^{2}r^{3}}+2e^{-2\phi}H\dfrac{\partial \phi}{\partial t}+\dfrac{(2r-b-rb')}{2a^{2}r^{2}}\dfrac{\partial \phi}{\partial r}+\dfrac{(r-b)}{a^{2}r}\left(\left(\frac{\partial \phi}{\partial r}\right)^{2}+\left(\frac{\partial^{2}\phi}{\partial r^{2}}\right)\right)=\kappa(p_{1}+p_{t}+\Pi)\label{eq7}
\end{eqnarray} and 
\begin{equation}
	2\dot{a}e^{-\phi}\left(\sqrt{\frac{r-b}{b}}\right)\dfrac{\partial \phi}{\partial r}=0\label{eq8}
\end{equation} Here, $\kappa=8\pi G$, $u^{\alpha}=(e^{-\phi},0,0,0)$ is the unit time-like vector, $H=\dfrac{\dot{a}}{a}$ is the usual Hubble parameter, an over dot indicates differentiation w.r.t $t$ while an over dash stands for radial differentiation. In the present work no radial energy flow is considered due to complexity in calculation. In particular, non-zero r.h.s of equation (\ref{eq8}) gives $r$- dependence of the red-shift function and as a result, the field equations become very complicated and it will be very hard to find any solution. As we are studying non-static WH solutions so from equation (\ref{eq8}) we have only one option namely $\dfrac{\partial \phi}{\partial r}=0$. So, in the following subsections we consider the followings $(i)~\phi=0$ and $(ii)~\phi=\phi(t)$. So far we have not restricted $b(r)$, the shape function of the WH configuration. But for a viable WH geometry the shape function $b(r)$ must satisfy the following restrictions \cite{Dutta:2023wfg}:
\begin{itemize}
	\item At the throat $(r=r_{0})$, the shape function should satisfy:
	$b(r_{0})=r_{0}$, $b'(r_{0})<1$.
	\item $b(r)<r$ for $r>r_{0}$.
	\item As red-shift function $\phi(r)$ is a representative of the horizons so for traversability there should not be any horizon at the throat (i.e, $\phi(r)$ should be finite at $r=r_{0}$) i.e, $((r-b(r))\phi'\rightarrow0)$ as $r\rightarrow r_{0}$.
	\item Flare out condition: A tenable geometry to the WH for possibility of human travel puts restrictions on the matter stress-energy tensor. Thus, the constraint of minimum radius at the throat $(r_{0})$ together with the traversability criterion results in tremendous tension at the throat. This appears as negative energy density, resulting in violation of null energy conditions at the throat. Mathematically, $\kappa(\rho+p_{r})<0$ which for the static WH gives
	$\dfrac{b'r-b}{r^{3}}+2(1-\frac{b}{r})\dfrac{\phi'}{r}<0$
\end{itemize}
\section{CWH configuration with zero red-shift function}
The Einstein field equations (\ref{eq4}-\ref{eq7}) of the previous section take the following simple form for zero red-shift function:
\begin{eqnarray}
	3H^{2}+\dfrac{3K}{a^{2}}+\dfrac{b'}{a^{2}r^{2}}=\kappa(\rho_{1}+\rho_{2})\label{eq9*}\\
	-(2\dot{H}+3H^{2}+\dfrac{K}{a^{2}})-\dfrac{b}{a^{2}r^{3}}=\kappa(p_{1}+p_{r}+\Pi)\label{eq9}\\
	-(2\dot{H}+3H^{2}+\dfrac{K}{a^{2}})+\dfrac{b-rb'}{2a^{2}r^{3}}=\kappa(p_{1}+p_{t}+\Pi)\label{eq10}
\end{eqnarray}
Now subtracting equation (\ref{eq9}) from equation (\ref{eq10}) gives
\begin{equation}
	p_{t}-p_{r}=\dfrac{(3b-rb')}{2a^{2}r^{3}}\label{eq11}
\end{equation}
Now using equation (\ref{eq11}) in the conservation equation (\ref{eq00}) and integrating we get,
\begin{equation}
	p_{r}=-\dfrac{b}{a^{2}r^{3}}+p_{r0}(t)\label{eq12}
\end{equation} with $p_{r0}(t)$ is an arbitrary integration function. A simple algebra of (\ref{eq11})+(\ref{eq12}) results in
\begin{equation}
	p_{t}=\dfrac{(b-rb')}{2a^{2}r^{3}}+p_{r0}(t)\label{eq13}
\end{equation} Now, if we assume $p_{t}=\alpha p_{r}$ with $(\alpha\neq1)$ being the anisotropy parameter then from (\ref{eq12}) and (\ref{eq13}) we get
\begin{equation}
	b(r)=p_{0}r^{3}+b_{0}r^{1+2\alpha}\label{eq17}
\end{equation} where $p_{r0}(t)=\frac{p_{0}}{a^{2}}$ and $p_{0}$, $b_{0}$ are constants of integration. Using the above expressions for $p_{r}$, $p_{t}$ and $b(r)$, the energy density $\rho_{2}$ can be integrated from the conservation equation (\ref{eq2*}) as
\begin{equation}
	\rho_{2}(r,t)=\dfrac{b_{0}(1+2\alpha)r^{2(\alpha-1)}}{a^{2}}+\dfrac{\rho_{0}(r)}{a^{3}}\label{eq18}
\end{equation} with $\rho_{0}(r)$ an arbitrary integration function. Also using $b(r)$ from equation (\ref{eq17}) to the pressure components in equation (\ref{eq12}) and (\ref{eq13}), one gets the simplified form of $p_{r}$ and $p_{t}$ as 
\begin{eqnarray}
	p_{r}=-\dfrac{b_{0}r^{2(\alpha-1)}}{a^{2}}\nonumber\\
	p_{t}=-\dfrac{\alpha b_{0}r^{2(\alpha-1)}}{a^{2}}\label{eq19**}
\end{eqnarray} Now to study the viability of the WH solution for which the shape function is given by (\ref{eq17}) we shall examine the conditions presented in the previous section. The condition $b(r_{0})=r_{0}$ eliminates the parameter $p_{0}$ to give 
\begin{equation}
	b(r)=r\left(\dfrac{r}{r_{0}}\right)^{3}\left(1+b_{0}r_{0}^{2\alpha}\left(\left(\dfrac{r}{r_{0}}\right)^{2\alpha-2}-1\right)\right)
\end{equation}
As in the present WH geometry, the redshift function $\phi$ is chosen to be zero so that the flare-out condition simplifies to be $\dfrac{b-rb'}{r^{3}}>0$. This restricts the radial coordinate $`r'$ as $r<\mu r_{0}$ with $\mu=\left(\dfrac{b_{0}r_{0}^{2\alpha}-1}{\alpha b_{0}r_{0}^{2\alpha}}\right)>1$ i.e,
\begin{equation} r_{0}>\left(\dfrac{1}{(1-\alpha)b_{0}}\right)^{\frac{1}{2\alpha}}\label{eq20}
	\end{equation} 
Thus, traversability is maintained within a finite range of the radial coordinate. Geometrically, the null or time-like trajectories are possible only within this finite range. In literature, such finite size WHs do exist where the traversability is ensured only within a finite range. One may refer to \cite{Cataldo:2017yec}, \cite{Ghosh:2021bmf}, \cite{Kim:1995xf} in this regard.
Now, using equations  (\ref{eq18}) and (\ref{eq19**}) into equations (\ref{eq9*})-(\ref{eq10}) we get
\begin{eqnarray}
	3H^{2}=\kappa \rho_{1}+\dfrac{\rho_{0}}{a^{3}}\label{eq21}\\
	-(2\dot{H}+3H^{2})=\kappa(p_{1}+\Pi)
\end{eqnarray} where $\rho_{0}(r)=\rho_{0}$ is chosen for consistency and $K=0$ is chosen for simplicity. Further, due to conservation equation (\ref{eq1*}) $\rho_{0}$ should be identically zero. Thus the equations (\ref{eq18}) and (\ref{eq21}) are simplified to
\begin{eqnarray}
	\rho_{2}(r,t)=\dfrac{b_{0}(1+2\alpha)r^{2(\alpha-1)}}{a^{2}}\label{eq19*}\\
	3H^{2}=\kappa \rho_{1}\label{eq20*}
	\end{eqnarray}
To determine the cosmic evolution we assume the thermodynamic pressure $p_{1}$ to be barotropic in nature i.e, $p_{1}=\omega \rho_{1}$ and the dissipative pressure component is chosen in the form
\begin{equation}
	\Pi=-\dfrac{\delta}{\rho_{1}^{n}}
\end{equation} Then the matter conservation equation (\ref{eq1*}) solves for $\rho_{1}$ to give
\begin{equation}
	\rho_{1}=
	\begin{cases}
		\left(\dfrac{\delta}{1+\omega}+\dfrac{z_{0}a^{-l}}{1+\omega}\right)^{\frac{1}{1+n}} & \text{for } \omega\neq-1\\
		3\delta(1+n)\ln\left(\frac{a}{a_{0}}\right)& \text{for } \omega=-1
	\end{cases}\label{eq26**}
\end{equation} where $z_{0}$, $a_{0}$ are integration constants.  For $n>0$, the dissipative pressure $\Pi$ approaches to zero as the energy density $\rho_{1}$ is very large while $\Pi$ becomes negative infinitely large for vanishing $\rho_{1}$.
\begin{figure}[h!]
\centering \includegraphics{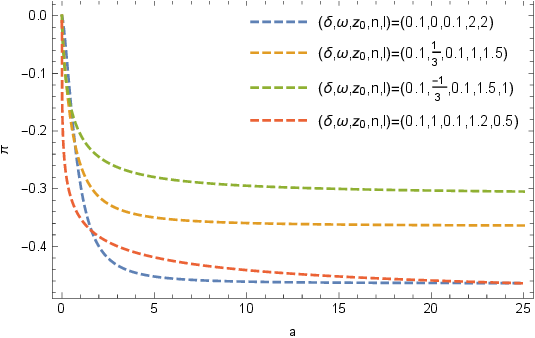}
\caption{Variation of dissipative pressure with cosmic scale factor for $\omega \neq -1$} \label{F2}
\end{figure}
\begin{figure}[h!]
	\centering \includegraphics{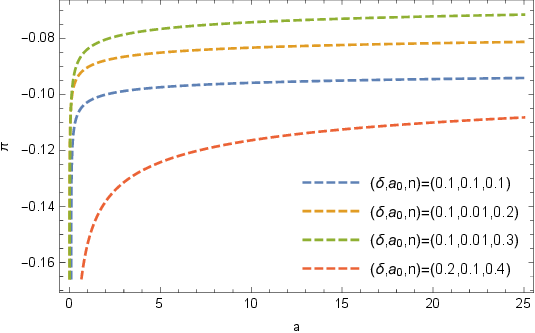}
	\caption{Variation of dissipative pressure with cosmic scale factor for $\omega=-1$} \label{F3}
\end{figure}
 The variation of $\Pi$, the dissipative pressure with scale factor $a$ for $\omega \neq -1$ has been presented in FIG. (\ref{F2}) which shows that the dissipative pressure sharply falls with the increase in scale factor in the early era and subsequently it becomes constant at large values of the scale factor. Further, the variation of dissipative pressure for $\omega=-1$ as presented in FIG. (\ref{F3}) shows that $\Pi$ sharply increases for small values of the scale factor and then becomes almost constant as the scale factor gradually increases. Now, using $\rho_{1}$ from equation (\ref{eq26**}) into equation (\ref{eq20*}) and assuming $\kappa=1$ we get the scale factor as
\begin{equation}
		\begin{cases}
	~~~~~~~~~~~~~~	\dfrac{\sqrt{3}}{2}(1+\omega)z_{0}^{\beta}(t-t_{0})=a^{\frac{\sqrt{3}}{2}(1+\omega)} 2^{F_{1}}\left(\beta,\beta,\beta+1,-\dfrac{\delta}{z_{0}(1+\omega)}a^{\frac{3(1+\omega)}{2\beta}}\right)& \text{for } \omega\neq-1\\
	~~~~~~~~~~~~~~~~~~~~~~~~~~~~~~~~~~~~~~~~~~~~	a=a_{0}\exp\left(\left(\dfrac{\sqrt{3}}{2}\delta(2n+1)\right)^{\frac{1}{1-\beta}}(t-t_{0})^{\frac{1}{1-\beta}}\right)& \text{for } \omega=-1
	\end{cases}
\end{equation} with $\beta=\dfrac{1}{2(n+1)}$ and $2^{F_{1}}$ is the usual Hyper-geometric function. It should be noted that if $-1<n<-\dfrac{1}{2}$ i.e, $\beta>1$ then $a\rightarrow a_{0}$ as $t\rightarrow\infty$. Thus the cosmic evolution of the WH configuration has an interesting feature of emergent scenario at infinite past for $\omega=-1$. 

We shall now examine the energy conditions for both the fluids separately. The explicit restrictions for the energy conditions are:
\begin{itemize}
	\item Null Energy Condition (NEC): $\rho+p_{r}\geq0$, $\rho+p_{t}\geq0$
	\item Weak Energy Condition (WEC): $\rho\geq0$, $\rho+p_{r}\geq0$, $\rho+p_{t}\geq0$
	\item Strong Energy Condition (SEC): $\rho+p_{r}\geq0$, $\rho+p_{t}\geq0$, $\rho+p_{r}+2p_{t}\geq0$
	\item Dominant Energy Condition (DEC): $\rho-|p_{r}|\geq0$, $\rho-|p_{t}|\geq0$
\end{itemize}
The restrictions for the energy conditions of the two fluids are
\begin{itemize}
	\item Fluid I: For $\alpha>0$ all the energy conditions hold.
	\item Fluid II: $\omega>-\dfrac{1}{3}$, $\delta>0$, $1+n>0$ satisfies all the energy conditions.
\end{itemize}
Based on the above analysis of the WH configuration in inhomogeneous space-time, the results can be summarized in the form of the following lemmas:

\begin{itemize}
	\item \textbf{\textit{Lemma-I:}} \textit{The WH configuration is fully characterized by the nature of anisotropic fluid. The anisotropy parameter has a crucial role in determining the nature of the WH.}
	\item \textbf{\textit{Lemma-II:}}\textit{The cosmic evolution does not depend on the nature of anisotropic fluid, it depends on the dissipative but homogeneous and isotropic fluid.}
	\item \textbf{\textit{Lemma-III:}}\textit{The cosmic evolution remains the same if the presently considered dissipative fluid is replaced by a barotropic fluid in the form of modified chaplygin gas. The result can be generalized as follows: The dissipative fluid can be replaced by a perfect fluid with appropriate barotropic equation of state.} 
\end{itemize}
	 It is to be noted that the present WH solution with zero red-shift function has some analogy with the work in \cite{Cataldo:2012pw} and \cite{Kim:1995xf}. But the basic difference is that here no equation of state is assumed rather a linearized anisotropy parameter is only considered. Contrary to \cite{Cataldo:2012pw}, one of the matter component namely, fluid II in our work depends both on $r$ and $t$.
	 Though, in the WH solution of \cite{Cataldo:2012pw} WEC is violated but in the present WH solution all the energy conditions are satisfied for both the fluids with some restrictions on the parameters involved.
\section{Role of red-shift function in the formation of CWH}
The present section analyzes the formation of CWH and the cosmic evolution in the background of inhomogeneous space-time.  The present section has some similarities with the work in \cite{Kim:1995xf}. In both the studies, it is examined whether traversability is possible with non-exotic matter or not. Also, the CWH solution in both the studies show that traversability is restricted within a finite region around the throat. From Einstein field equations (\ref{eq8}), for non-static geometric consideration the red-shift function is $r$-independent so we choose $\phi=\phi(t)$, to examine its role in WH geometry as well as in the evolution of the scale factor.  One can identify the red-shift function as potential of the WH. It is important to note that for purely temporal dependence of the red-shift function one may define the time coordinate as
\begin{equation}
	T=\int \exp(\phi)dt.
\end{equation} This may eliminate the effect of the tidal force (i.e, the red-shift function) but simultaneously, the scale factor $a$ has been changed. As a result, the WH solution is also changed. Now, due to $\dfrac{\partial \phi}{\partial r}=0$ from equation (\ref{eq8}), the field equations (\ref{eq4}) to (\ref{eq7}) reduces to
\begin{eqnarray}
	3e^{-2\phi(t)}H^{2}+\dfrac{b'}{a^{2}r^{2}}+\dfrac{3K}{a^{2}}=\kappa(\rho_{1}+\rho_{2})\label{eq28}\\
	-e^{-2\phi(t)}(2\dot{H}+3H^{2})+\dfrac{K}{a^{2}}-\dfrac{b}{a^{2}r^{3}}+2e^{-2\phi}H\dfrac{\partial \phi}{\partial t}=\kappa(p_{1}+p_{r})\label{eq29}\\
	-e^{-2\phi(t)}(2\dot{H}+3H^{2})+\dfrac{K}{a^{2}}+\dfrac{b-rb'}{2a^{2}r^{3}}+2e^{-2\phi}H\dfrac{\partial \phi}{\partial t}=\kappa(p_{1}+p_{t})\label{eq30}
\end{eqnarray} while the conservation equations (\ref{eq1*})-(\ref{eq00}) remain the same. As in the previous section if we consider the difference between anisotropic pressure components from the field equations (\ref{eq29}) and (\ref{eq30}) we obtain the same expression as in equation (\ref{eq11}). As a result, using the conservation equation (\ref{eq00}) (i.e, the relativistic Euler equation) we have the identical expressions for the radial and transverse pressure components for Fluid II (as given in equations (\ref{eq12}) and (\ref{eq13})). Similarly, the shape function and the energy density for the second fluid remain identical due to linear relation between the anisotropic pressure components and the conservation equation (\ref{eq2*}). Now using the components for Fluid II and the shape function in the above field equations (\ref{eq28})-(\ref{eq30}) we have two simple forms of evolution equations (choosing $\kappa=1$) as 
\begin{eqnarray}
	3H^{2}=\rho_{1}e^{2\phi}\label{eq32}\\
	-2\dot{H}-3H^{2}+2\dot{\phi}H=\kappa p_{1}e^{2\phi}
\end{eqnarray} with conservation equation (\ref{eq1*}). Now choosing barotropic equation of state i.e, $p_{1}=\omega \rho_{1}$, one gets the evolution equation as
\begin{equation}
	2\dot{H}+3(1+\omega)H^{2}=2\dot{\phi}H\label{eq33}
\end{equation} We now consider two cases for explicit cosmic evolution.\\
Case-I: $\omega=\omega_{0}$, a constant.\\
The equation (\ref{eq33}) has a first integral as
\begin{equation}
	e^{2\phi}=\phi_{0}H^{2}a^{3(1+\omega_{0})}\label{eq34}
\end{equation} with $\phi_{0}$, being the constant of integration. The conservation equation gives
\begin{equation}
	\rho=\rho_{0}a^{-3(1+\omega_{0})}\label{35},
\end{equation} $\rho_{0}$ is the integration constant. For an explicit form of the scale factor the red-shift function $\phi$ is assumed to be in power law form as
\begin{equation}
	e^{\phi}=\delta_{0}t^{n}
\end{equation} $\delta_{0}$, $n$ are constants so that equation (\ref{eq34}) has the solution 
\begin{equation}
	a(t)=
	\begin{cases}
		a_{0}t^{m},~m=\dfrac{2(n+1)}{1+\omega_{0}} & \text{for } \omega_{0}\neq-1\\
	a_{0}\exp(lt)& \text{for } \omega_{0}=-1
	\end{cases}\label{eq26}
\end{equation} 
It is to be noted that all the energy conditions are satisfied if $\omega_{0}>-\dfrac{1}{3}$.\\
Case-II Modified Chaplygin Gas:\\
We have
\begin{equation}
	p=(\gamma-1)\rho-\dfrac{\delta}{\rho^{n}}\label{eq38}
\end{equation} so that
\begin{equation}
	\omega=\dfrac{p}{\rho}=(\gamma-1)-\dfrac{\delta}{\rho^{n+1}}\label{eq39}
\end{equation}
 Uisng (\ref{eq38}) in the conservation equation (\ref{eq1*}) gives
 \begin{equation}
 	\rho=\left(\dfrac{\delta}{\gamma}+\dfrac{\rho_{0}}{\gamma}a^{-3\gamma(n+1)}\right)^{\frac{1}{n+1}}\label{eq40}
 \end{equation} Using (\ref{eq39}) and (\ref{eq40}) in the evolution equation (\ref{eq33}) one gets after integration,
\begin{equation}
	e^{\phi}=\phi_{0}Ha^{\frac{3}{2}\rho_{0}\alpha}\left(\rho_{0}+\delta a^{3\mu}\right)^{-\left(\frac{\rho_{0}\alpha}{2\mu}\right)}
\end{equation} with $\mu=\gamma(n+1)$. Now choosing $e^{\phi}=l_{0}t^{s}$, the explicit form of the scale factor is given by
\begin{equation}
	a^{\frac{3\rho_{0}\alpha}{2}}2^{F_{1}}\left(\dfrac{\rho_{0}\alpha}{2\mu},\dfrac{\rho_{0}\alpha}{2\mu};1+\dfrac{\rho_{0}\alpha}{2\mu};-\dfrac{\delta}{\rho_{0}}a^{3\mu}\right)=a_{0}t^{s+1}
\end{equation} where $a_{0}=\dfrac{3\rho_{0}\alpha l_{0}}{2\phi_{0}(s+1)}\rho_{0}^{\frac{\rho_{0}\alpha}{2\mu}}$, $\mu=\alpha(n+1)$ and $2^{F_{1}}$ is the usual Gauss hyper-geometric function.\\

\begin{figure}[h!]
	\begin{minipage}{0.3\textwidth}
	~~~~~~~~~~~~~~~~~~~~~~~~~~~~~~~~~~~~~~~~~~~~~~~~~~~~~	\centering\includegraphics[height=6cm,width=9cm]{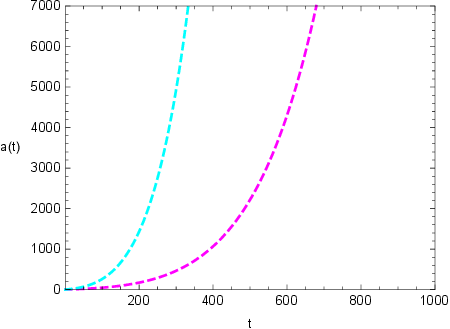}
	\end{minipage}
	\caption{[$a(t)$ vs $t$ for (i) $\rho_0=1,~\alpha=1,~l_{0}=0.01,~\phi_{0}=0.01,~s=-0.95,~n=0.06,~\beta=0.1,~\mu=1.06$ (Dashed cyan); (ii) $\rho_0=1,~\alpha=1,~l_{0}=0.01,~\phi_{0}=0.01,~s=-0.5,~n=0.05,~\beta=0.02,~\mu=1.05$ ( Dashed Magenta)]}\label{f01}
\end{figure}
The graphical representation of the scale factor vs cosmic time is shown in FIG-\ref{f01} for two sets of choices of the parameters involved.
Further using the property of hyper-geometric functions namely,
\begin{equation}
	x2^{F_{1}}(1,1,2,-x)=\ln(1+x)
\end{equation} and  one may get an emergent scale factor given by
\begin{equation}
	a(t)=\left(\dfrac{\rho_{0}\alpha}{2\mu}-\exp\left(a_{0}t^{s+1}\right)\right),~\dfrac{\rho_{0}\alpha}{2\mu}=1
\end{equation} since  $a\rightarrow \dfrac{\rho_{0}\alpha}{2\mu}$ (constant) as $t\rightarrow-\infty$. In cosmological context, the above solution for the scale factor shows that the present model indicates a contracting model of the universe having big crunch singularity at $t=0$ and hence it is not of much interest. However, from the point of view of the energy conditions, it is easy to see that if $\rho_{0}>0$, $\dfrac{2}{3}<\gamma<2$ and $\delta>0$, then all the energy conditions are satisfied except the SEC $(\rho+3p\geq0)$ which is indefinite in sign.  In any case if, $\delta>0$ is very small then it is definitely satisfied.  The solution obtained in the present work is different from that given in \cite{Kim:1995xf} due to differences in physical assumptions.
The above results can be summarized in the form of a Lemma as follows:\\

\begin{itemize}
	\item \textbf{\textit{Lemma-IV:}} \textit{The inhomogeneous and anisotropic fluid (i.e, Fluid II) as well as the shape function do not depend on the red-shift function. Hence, WH configuration has no effect of the gravitational tidal force (characterized by the red-shift function). The red-shift function only influences the cosmic evolution. This is a distinct feature in CWH scenario compared to the static WH and the $(r-t)$ equation (\ref{eq8}) has a crucial role for identifying $\phi_{r}=0$}
\end{itemize}
\section{Summary and concluding remarks}
CWH prescription has been discussed from an analytical point of view for inhomogeneous FLRW space-time geometry. Assuming linearity relation among the pressure components of the inhomogeneous and anisotropic fluid, shape function has been evaluated and it is found that the anisotropy parameter has distinct role both in characterization of the WH geometry as well as in the fluid components. However, $\alpha$ (anisotropy parameter) does not influence the homogeneous fluid and the cosmic evolution. It is interesting to note that except linearity no other choices for the interrelation among the inhomogeneous and anisotropic fluid pressure components is permissible due to non-static nature of the space-time geometry. Hence, one may claim that the present WH configuration is the unique WH solution for the inhomogeneous FLRW model. Also, it is interesting to note that choosing $a(t)=a_{0}$, it is possible to have the static WH geometry. Moreover, an important point to be noted is that instead of the present two fluid system, it is not possible to have a consistent WH geometry with a single inhomogeneous and anisotropic fluid in non-static space-time geometry. The energy conditions are examined for both the fluids separately. For Fluid I, all the energy conditions are satisfied provided the anisotropy parameter, $\alpha$ is positive while in case of Fluid II all the energy conditions hold if ($\omega>-\dfrac{1}{3}$ (i.e, usual fluid), $\delta>0$ and $n>-1$). Thus, in the present physical scenario the two fluids being considered have two distinct physical roles. Fluid I (homogeneous and isotropic but dissipative in nature) does not influence the WH geometry, it only indicates the cosmic evolution of the background geometry. On the other hand, Fluid II (inhomogeneous and anisotropic) has significant role in forming the WH configuration via the anisotropy parameter $\alpha$ and it does not affect the cosmic evolution. Finally, for future work it will be interesting to consider WH geometry for anisotropic fluid having heat flow component to examine the role of red-shift parameter prominently.
	\section*{Acknowledgment}
 The authors thank the anonymous referees for their valuable and insightful comments that improved the quality and visibility of the work. M.C thanks University Grants Commission (UGC) for providing the Senior Research Fellowship (ID:211610035684/JOINT CSIR-UGC NET JUNE-2021). S.C thanks FIST program of DST, Department of Mathematics, JU (SR/FST/MS-II/2021/101(C)). The authors are thankful to Inter University Centre for Astronomy and Astrophysics (IUCAA), Pune, India for their warm hospitality and research facilities where this work was carried out under the ``Visiting Research Associates" program of S.C.

\end{document}